\documentclass[letterpaper,11pt]{article}
\usepackage{jheppub}
\usepackage{amsmath}
\usepackage{physics}
\usepackage{relsize}
\usepackage{tikz}
\usepackage{pgfplots}
\usepackage{comment}

\usepackage[utf8]{inputenc}

\usepackage{bbm}
\newcommand{\cev}[1]{\reflectbox{\ensuremath{\vec{\reflectbox{\ensuremath{#1}}}}}}
\newcommand{\p}{\boldsymbol{p}}
\newcommand{\x}{\boldsymbol{x}}

\title{$T\bar{T}$ deformed scattering happens within matrices}

\author[a]{Vasudev Shyam}
\author[b]{Yigit Yargic}

\affiliation[a]{Stanford Institute for Theoretical Physics and Department of Physics, Stanford University, Stanford, CA 94305, USA}
\affiliation[b]{Microsoft Research, Redmond WA 98052, USA}
\emailAdd{vshyam@stanford.edu}
\emailAdd{v-yyargic@microsoft.com}

\begin{document}

\abstract{\\
We show the $T\bar{T}$ deformation of two-dimensional quantum field theories is equivalent to replacing the spacetime dependence of the fields with dependence on the indices of infinitely large matrices. We show how this correspondence explains the CDD phase dressing of the $S$-matrix and the general formula for the deformation of arbitrary correlation functions. We also describe how the Moyal deformation of self-dual gravity is a $T\bar{T}$ deformation of the theory described by the Chalmers-Siegel action, where the $T\bar{T}$ deformation is defined on the two-dimensional plane of interactions.}

\maketitle

\section{Introduction}

The irrelevant $T\bar{T}$ deformation \cite{Smirnov:2016lqw,Cavaglia:2016oda} of two-dimensional quantum field theories has proven to be a very exciting development in modern quantum field theory.
It presents a peculiar puzzle where despite it being irrelevant from the perspective of the renormalization group, various observables of a given theory deformed by $T\bar{T}$ can be exactly computed as functions of the deformation parameter. The locality properties of the underlying un-deformed theories however are drastically modified. It was first discovered in the context of integrable models, where it was recognized as the first in a tower of deforming operators formed from a quadratic combination of conserved currents that preserve integrability. Since it is a quadratic combination of the energy-momentum tensor, which any relativistic (or Euclidean invariant) two dimensional field theory possesses, the $T\bar{T}$ deformation of any two-dimensional quantum field theory can be studied. 

Perspectives about what the effect of the deformation ``truly" is range from seeing the deformation as a coupling of a quantum field theory to random geometry \cite{Cardy:2018sdv}, flat space Jackiw-Teitelboim gravity \cite{Dubovsky:2017cnj, Dubovsky:2018bmo}, which was later recognized to be a rewriting of ghost-free massive gravity \cite{Tolley:2019nmm, Freidel:2008sh, Mazenc:2019cfg}, all the way to performing a field-dependent change of coordinates of the underlying theory \cite{Conti:2018tca}, replacing point like particles with rigid rods \cite{Cardy:2022olv}, and imbuing a material described by the undeformed theory with exotic elasticity \cite{Cardy:2022mhn}. It was also recognized in \cite{McGough:2016lol} that the $T\bar{T}$ deformation of holographic CFTs leads to a theory whose holographic dual is gravity in AdS$_3$ but with a Dirichlet wall at finite radius.  

In the following article, we provide a simpler alternative representation of the $T\bar{T}$ deformation of a given two-dimensional quantum field theory. We propose that the effect of the deformation is the same as replacing the spacetime dependence of the fields with dependence on the indices of an infinite-sized matrix. In fact, the relationship between Moyal deformed operator algebras and the $T\bar{T}$ deformation has been anticipated already in \cite{Lechner08, Grosse:2008dk, Buchholz:2010ct, Lechner:talk}. The novelty of our approach is that we do not see the Moyal $\star$-product as arising from underlying spacetime non-commutativity, but as a dressing of the field interactions on a commutative spacetime. We present an explicit prescription based on a simple formula for the components within the product of matrices which reproduces the effects of the deformation.

Our geometrization of matrix indices is adapted from the Wigner-Weyl transformation of quantum operators \cite{Zachos2017, Banburski:2022gcj}. In any matrix representation, two combinations of the row and column indices behave as conserved charges during the matrix multiplication: their difference, and the Fourier conjugate of their sum. These combinations span a commutative momentum space on which the matrix entries live. As two matrices are multiplied, the entries of their product are given by a $\star$-convolution, rather than an ordinary convolution. This is the origin of non-locality in matrix models, which we understand here as $T\bar{T}$-deformations. In other words, the geometry of $T\bar{T}$-deformed theories is the geometry of matrix indices!

\subsection*{Organization of the paper}

In Section 2 we describe the basic construction of how geometry is seen from a re-interpretation of certain combinations of matrix indices, and how a $\star$-product structure naturally emerges from the matrix product. 

Section 3 shows how this insight applied to $S$-matrices of two-dimensional quantum field theories allows us to recover the Castillejo-Dalitz-Dyson (CDD) phases associated to the $T\bar{T}$ deformation. 

Section 4 describes how the deformation of general correlation functions in arbitrary Euclidean quantum field theories also follows from the replacement of fields by their matrix counterparts. 

Finally, Section 5 explains how our prescription applies at the level of the action of the theory. We do a case study of the the Moyal deformation of self-dual gravity and explain how it is in fact a $T\bar{T}$ deformation. We then highlight the relationship between the natural Heisenberg algebra inherent to our matrix formalism and the $W_{\wedge}$ algebra that features in the study of the Moyal deformation of self-dual gravity.

\section{Matrix geometry}
\label{sec:MatGeo}

Let $Q$ and $\tilde{Q}$ be two matrices which satisfy $[Q,\tilde{Q}] = \tfrac{i}{2\kappa} \mathbbm{1}$, where $\kappa$ is a constant. We will use these matrices to parametrize the entries of any matrix with commuting arguments.

We start by noting that the $Q$-commutator and $\tilde{Q}$-commutator maps commute with each other due to the Jacobi identity,
\begin{align}
\label{comcom1}
    [Q, [\tilde{Q}, \bullet]] - [\tilde{Q}, [Q, \bullet]] = [[Q, \tilde{Q}], \bullet] = 0
    .
\end{align}
Therefore, the space spanned by the eigenvalues of $Q$-commutator and $\tilde{Q}$-commutator is commutative. Similarly, the $Q$-anticommutator and $\tilde{Q}$-anticommutator also commute with each other,
\begin{align}
\label{comcom2}
    \{Q, \{\tilde{Q}, \bullet\}\} - \{\tilde{Q}, \{Q, \bullet\}\} = [[Q, \tilde{Q}], \bullet] = 0
    ,
\end{align}
therefore their eigenvalues span another commutative space.

We denote the aforementioned eigenvalues by $p, \tilde{p}, \tilde{x}, x \in \mathbb{R}$. Acting on an arbitrary matrix $A$, these maps have the eigenvalue equations
\begin{align}
\label{comcom3}
\begin{array}{lll}
    \displaystyle
    [Q, A](p) = p A(p)
    & , \qquad &
    \displaystyle
    \{Q, A\}(\tilde{x}) = \tfrac{1}{\kappa} \tilde{x} A(\tilde{x})
    \;, \\
    \displaystyle
    [\tilde{Q}, A](\tilde{p}) = \tilde{p} A(\tilde{p})
    & , \qquad &
    \displaystyle
    \{\tilde{Q}, A\}(x) = - \tfrac{1}{\kappa} x A(x)
    \;.
\end{array}
\end{align}
The relations \eqref{comcom1} and \eqref{comcom2} imply two commutative sets of these parameters,
\begin{align}
    [p,\tilde{p}] = 0
    , \qquad
    [x, \tilde{x}] = 0
    .
\end{align}
We identify $(p,\tilde{p})$ as the coordinates of a commutative momentum space, and $(x,\tilde{x})$ as the coordinates of a commutative position space.

The definitions \eqref{comcom3} are consistent with the Heisenberg uncertainty principle. From $\{-\kappa \tilde{Q}, [Q, \bullet]\} - [Q, \{-\kappa\tilde{Q}, \bullet\}] = \kappa \{[Q, \tilde{Q}],\bullet\} = i \bullet$, we get $[x, p] = i$, and similarly $[\tilde{x}, \tilde{p}] = i$.

To be explicit, we represent all matrices in the eigenbasis of $Q$, such that for continuous row and column indices $a,b \in \mathbb{R}$, we write $Q{}_{ab} = a \delta(a-b)$. In this basis, the parameter $p = a - b$ is the index difference, while the parameter $\tilde{x} = \kappa (a+b)$ is the index sum.

Furthermore, the doublets $(x,\tilde{x})$ and $(p, \tilde{p})$ are Fourier conjugate to each other. To see this, we recall from quantum mechanics that the eigenvalues of the left-multiplication by $Q$ and by $\tilde{Q}$ are related by a Fourier transformation, therefore there is a unitary matrix $U$ with the entries $U_{ab} \propto e^{2i\kappa a b}$
such that $\tilde{Q} = U Q U^\dagger$. Then,
\begin{align}
    (U^\dagger A U)_{ab} 
    &= \sum_{c, d} e^{-2i\kappa (ac-bd)} A_{cd}
    \nonumber \\
    &= \sum_{c, d} e^{-i \kappa (a+b)(c-d)} e^{i \kappa (c+d)(a-b)} A_{cd}
\end{align}
shows that $x = -\kappa (a+b)$ and $p = c-d$ are Fourier conjugate to each other, as well as $\tilde{x} = \kappa (c+d)$ and $\tilde{p} = a-b$.

The matrices $Q$ and $\tilde{Q}$ serve here only to fix a basis for this geometric parametrization of matrix entries. To summarize, we can start with any model of matrices with no prior geometry, represented in an arbitrary but fixed basis. Then, for each matrix entry at the row and column labels $(a,b) \in \mathbb{R}^2$, we identify the index difference $p = a-b$ as a momentum parameter in the first direction, and the index sum $\tilde{x} = \kappa (a+b)$ as a position parameter in the second direction. The parameters $x$ and $\tilde{p}$ are then defined through Fourier transformation.

We now check how the matrix interactions are manifested on this geometry. Given two matrices $A$ and $B$, we consider their product
\begin{align}
\label{Moyal0}
    (AB)_{ab} = \sum_c A_{ac} B_{cb}
    .
\end{align}
We will exchange the row and column indices $a,b,c$ in \eqref{Moyal0} by $\tilde{x} = \kappa (a+b)$, $p = a-b$, and $p' = a-c$. These relations can be reversed to $\kappa(a+c) = \tilde{x} + \kappa (p-p')$ and $\kappa (c+b) = \tilde{x} - \kappa p'$. Therefore,
\begin{align}
\label{Moyal1}
    (A \star B)(p, \tilde{x}) = \int \dd p' \, A(p', \tilde{x} + \kappa (p-p')) \, B(p-p', \tilde{x} - \kappa p')
    ,
\end{align}
where we used $\star$ to emphasize that this operation is still equivalent to matrix multiplication. We note that the momentum $p$ is conserved under the matrix multiplication, in the sense that the total momentum is the same for each term on both sides in \eqref{Moyal1}. This momentum conservation is a consequence of the Leibniz rule $[Q,AB] = [Q,A]B + A[Q,B]$.

We Fourier transform the index sum $\tilde{x}$ to $\tilde{p}$ in \eqref{Moyal1}, and obtain
\begin{align}
\label{Moyal2}
    (A \star B)(p, \tilde{p}) = \int \dd p' \dd \tilde{p}' \, e^{i \kappa (\tilde{p}' p - p' \tilde{p})} \, A(p', \tilde{p}') \, B(p-p', \tilde{p} - \tilde{p}')
    .
\end{align}
We note that the Fourier conjugate of the matrix index sum, $\tilde{p}$, is another conserved charge of the matrix multiplication. Since it commutes with $p$, writing the matrices as functions of $(p,\tilde{p})$ is justified.

The operation \eqref{Moyal2} is known as Moyal's $\star$-convolution \cite{Zachos2017}. The exponential factor $e^{i \kappa (\tilde{p}' p - p' \tilde{p})}$, which we will refer to as the \emph{Moyal factor}, distinguishes this expression from an ordinary convolution, and it captures the essence of how a matrix algebra differs from an algebra of local functions.

The Moyal factor can also be understood as making a non-local deformation of the field interactions by the parameter $\kappa$, where the locality would be recovered at $\kappa = 0$ \cite{Banburski:2022gcj}. The non-locality can be seen in \eqref{Moyal1} and in the similar expression written in terms of $(x, \tilde{p})$. As we explain next, these deformations of the field interactions along different values of $\kappa$ are equivalent to the $T\bar{T}$ deformations in a quantum field theory.

\section{\texorpdfstring{The $S$-matrix of matrix fields}{The S-matrix of matrix fields}}

The $S$-matrix evolves scattering states from the far past to the far future,
\begin{equation}
    S\vert \p_{1}\cdots \p_{l}\rangle_{\mathrm{in}} = \vert \p_{l}\cdots \p_{n}\rangle_{\mathrm{out}}.
\end{equation}
We study 2D unitary theories respecting crossing symmetry. 
We write the two-momentum of the $i$-th particle as 
\begin{equation}
    \p_i = (p_i, \tilde{p}_i).
\end{equation}
The LSZ reduction formula implies we can write the $S$-matrix in terms of expectation values of time-ordered products of field operators,
\begin{equation}
    S_{n}(\p_1, \ldots, \p_n) = \prod_{i}(\p^2+m^2)_{i} \, \langle\Omega\vert\mathcal{T}\bigg(\varphi(\p_1)\cdots\varphi(\p_n)\bigg)\vert \Omega\rangle\bigg\vert_{\p^2_{i}+m_i^2=0,\,\,\forall i.}
\end{equation}
In two dimensions it is customary to fix the momentum ordering. In terms of rapidities,
\begin{equation}
    p_i = m\cosh \theta_i
    ,\quad
    \tilde{p}_i = m\sinh \theta_i
    .
\end{equation}
this ordering implies that we take $\theta_1> \theta_2>\cdots > \theta_n$. 

We will now entertain the case where the momentum dependence of the fields is replaced by dependence on combinations of matrix indices as described in Section \ref{sec:MatGeo}. Then, from the momentum space perspective, the regular time-ordered products are replaced by the $\star$-products involving Moyal factors:
\begin{align}
    &
    \prod_{i}(\p^2+m^2)_{i} \, \langle \Omega \vert\mathcal{T} \bigg( \varphi(\p_1) \cdots \varphi(\p_n) \bigg) \vert \Omega\rangle \bigg\vert_{\p^2_{i}+m_i^2=0,\,\,\forall i}
    \nonumber \\
    \longrightarrow \; &
    \prod_{i}(\p^2+m^2)_{i} \, \langle \Omega \vert\mathcal{T} \bigg( \varphi(\p_1) \star \cdots \star \varphi(\p_n) \bigg) \vert \Omega \rangle \bigg\vert_{\p^2_{i}+m_i^2=0,\,\,\forall i}
    \nonumber \\ & \;
    = \exp\!\left(i\kappa \sum^{l}_{i<j=1} \p_{i}\times \p_{j}\right)
    \exp\!\left(-i\kappa \sum^n_{i<j=l} \p_{i}\times \p_{j}\right)
    \nonumber \\ & \qquad \cdot
    \prod_{i}(\p^2+m^2)_{i} \, \langle\Omega\vert \mathcal{T} \bigg( \varphi(\p_1) \cdots \varphi(\p_n) \bigg) \vert\Omega\rangle \bigg\vert_{\p^2_{i}+m_i^2=0,\,\,\forall i}
\end{align}
where
\begin{equation}
    \p_{i}\times \p_{j} \equiv \p^{\mu}_{i} \p^{\nu}_{j} \epsilon_{\mu\nu}
    = \tilde{p}_i p_j - p_i \tilde{p}_j
    .
\end{equation}
In summary, we can summarize the above relation as 
\begin{equation}
    S^{\star}_{n}(\p_1,\cdots\!, \p_n) = \exp\!\left(i \kappa \sum^{l}_{i<j=1} \p_{i}\times \p_{j}\right)\exp\!\left(-i\kappa \sum^n_{i<j=l} \p_{i}\times \p_{j}\right) S_n(\p_1,\cdots\!, \p_n)
    .
\end{equation}
In the next section, we will interpret the meaning of this phase. For now, we write these phases in the rapidity parameterization. We take
\begin{equation}
     \p_{i} = m(\cosh(\theta_i),\sinh(\theta_i))
     .
\end{equation}
Then we find that the terms in the exponent can be written as
\begin{equation}
    \exp\!\bigg(i\kappa \sum_{i<j} \p_{i}\times \p_{j}\bigg) = \exp\!\bigg(i\kappa m^2\sum_{i<j}\sinh(\theta_{ij})\bigg)
\end{equation}
where $\theta_{ij}=\theta_i - \theta_j$.

\subsection{\texorpdfstring{Recognizing the $T\bar{T}$ deformation}{Recognizing the TTbar deformation}}

So far we have shown that when we trade the space dependence of our fields with matrix index dependence, the $S$-matrix gets dressed by a phase. We note here that the phase of interest is in fact the form factor of the famous $T\bar{T}$ operator. In particular this operator is defined as
\begin{equation}
    T\bar{T} := T^{\mu\nu}T_{\mu\nu}-(T^{\rho}_{\rho})^2 = -\frac{1}{2} \det\!\left( T^{\alpha\beta}\right).
\end{equation}
The matrix elements of this operator in the scattering states is given by
\begin{equation}
    {}_{\mathrm{out}\!}\langle \p_{n} \cdots \p_l\vert \kappa(T\bar{T})\vert \p_{l}\cdots \p_1\rangle_{\mathrm{in}} = \left(-i\kappa \sum^l_{s<r=1} (\p_s \times \p_r) + i\kappa \sum^n_{i<j=l} (\p_i\times \p_j) \right).
\end{equation}
The dressing of the $S$-matrix is the result of performing the $T\bar{T}$ deformation of the original theory with a finite deformation parameter $\kappa$!

This fact was shown first in the context of integrable models in \cite{Cavaglia:2016oda}, then generalized to non-integrable theories in \cite{Dubovsky:2017cnj}.

Specializing to integrable theories (following \cite{Frolov:2019nrr} for instance), we deduce the deformation of the total energy due to $T\bar{T}$,
\begin{equation}
    \mathcal{E} = \sum^n_{j=1}\tilde{p}_j\rightarrow \mathcal{E}_{\kappa}
    ,
\end{equation}
via the Bethe equations
\begin{equation}
   e^{ip_i R} \prod_{l\neq i}S_2(\p_i, \p_l) =1.\label{}
\end{equation}
Here $R$ is the radius of the spatial circle (taken to be large), and $S_2(\p_i,\p_l)$ is the elastic two-body $S$-matrix element of the integrable theory. 

The dressing under $T\bar{T}$, or alternatively the procedure of replacing the space dependence of the fields by dependence on matrix indices, leads to the modified Bethe equation
\begin{equation}
     e^{ip_i R} \prod_{l\neq i}e^{i\kappa \left(\tilde{p}_ip_l-\tilde{p}_lp_i\right)} S_2(\p_i,\p_l) = 1.
\end{equation}
This can be recast in the form of the original Bethe equation but with a modification of $R$ to $\hat{R}$, i.e.,
\begin{equation}
    e^{ip_i R} \prod_{l\neq i}e^{i\kappa \left(\tilde{p}_ip_l-\tilde{p}_lp_i\right)}S_2(\p_i,\p_l) = e^{ip_i\hat{R}} \prod_{l\neq i}S_2(\p_i,\p_l)
\end{equation}
where 
\begin{equation}
    \hat{R} = R+\kappa \mathcal{E}
    .
\end{equation}
Then, the deformed energy is given by
\begin{equation}
    \mathcal{E}_{\kappa}(R) =\mathcal{E}(R-\kappa \mathcal{E}).
\end{equation}
This means the deformed energy $\mathcal{E}_{\kappa}$ is a solution to the inviscid Burgers' equation,
\begin{equation}
    \partial_{\kappa}\mathcal{E}_{\kappa}(R) = \mathcal{E}_{\kappa}(R)\partial_{R}\mathcal{E}_{\kappa}(R).
\end{equation}
This is a hallmark result for how the energy spectrum of a given theory flows under deformation by $T\bar{T}$, featured in \cite{Smirnov:2016lqw, Cavaglia:2016oda}.

If we look at the case with non-zero total momentum $J$, similar arguments to those above show that the flow equation for the energy is
\begin{equation}
    \partial_{\kappa}\mathcal{E}_{\kappa}(R) = \mathcal{E}_{\kappa}(R)\partial_{R}\mathcal{E}_{\kappa}(R)-J^2/R.
\end{equation}

\subsection{Free fields scatter non-trivially within matrices}
Let us now look at the case where we compute the $S$-matrix via the LSZ reduction formula for free fields, where we take all momenta to be incoming for simplicity,
\begin{align}
    {}_{\mathrm{out}\!}\langle 0  \vert S^{f\star} \vert \p_{1}\cdots \p_n \rangle_{\mathrm{in}}
    &\equiv S^{f\star}_{n}(\p_1\cdots \p_n)
    \\
    &= \prod_{i}\left(\p^2_i + m^2\right) \langle \Omega\vert \mathcal{T}\bigg(\phi(\p_n)\star\cdots \star\phi(\p_1)\bigg)\vert \Omega\rangle
    .
\end{align}
Here, to specify that the fields are free, we denoted them as $\phi(\p_i)$ which have a non-trivial two-point function
\begin{equation}
    \langle \Omega \vert \mathcal{T}(\phi(-\p) \phi(\p))\vert \Omega\rangle = G_{F}(\p)
    ,
\end{equation}
where the right-hand side is the Feynman propagator, the formal inverse of the kinetic operators appearing in the $S$-matrix formula. 

All higher-point correlators are given in terms of this two-point correlator by Wick's theorem. This is usually enough to show that all $S$-matrix elements are products of multiple kinetic operators and Green's functions, leaving us with $S=1$ as a result. This is not immediately true for the matrix version of the calculation laid out here. Instead we first have to filter out the Moyal factors by expressing the $\star$-product in terms of the regular product, so as to leave us finally with only the dressing,
\begin{align}
\label{eq:NG}
    \prod_{i} (\p^2_i+m^2) \, \langle \Omega\vert \mathcal{T}\bigg(\phi(\p_n)\star\cdots \star\phi(\p_1)\bigg)\vert \Omega\rangle
    = \exp\!\bigg(\!-\!i\kappa \sum^{n}_{i<j=1} \p_i \times \p_j\bigg)\cdot 1
    .
\end{align}
Note that $n$ has to be an even integer or else the $S$-matrix element vanishes. As a result, the above is a product of some power $l = n/2$ of two-body phases,
\begin{equation}
    \exp\!\bigg(\!-\!i\kappa \sum^{n}_{i<j=1}\p_i\times \p_j\bigg)
    = \prod_{j<i=1} \exp\!\bigg(\!-\!i\kappa \, \p_i\times \p_j\bigg),
\end{equation}
or
\begin{equation}
    S^{f\star}_{n=2l}(\p_1\cdots \p_{n}) = \prod^n_{i<j=1} S^{f\star}_{2}(\p_i,\p_j).
\end{equation}
This is a signature of an integrable theory. Furthermore, if we put the phases into the rapidity basis,
\begin{equation}
    S^{f\star}_{2}(\theta_1,\theta_2) = \exp\!\big(\!-\!m^2\kappa \sinh(\theta_{12})\big)
\end{equation}
and then take a scaled limit of the mass to zero as we perform an ultra-relativistic boost of the particles in opposite directions, we obtain the phase
\begin{equation}
    S^{f\star}_{2}(\theta_1,\theta_2)\vert_{m\rightarrow 0} = \exp\!\big(\!-\!\kappa \, p_{L,1} \, p_{R,2}\big)
    ,
\end{equation}
where $p_{L},p_{R}$ are the light cone momenta of the particles. This is the phase shift experienced by excitations scattering along an infinitely long, free bosonic string in Minkowski space! \cite{Dubovsky:2012wk}

\section{General correlation functions}

The phases in the previous section arose as a general consequence of the spacetime dependence of two-dimensional fields being replaced by matrix index dependence. This begs the question of what happens when we make this replacement within general correlation functions of two-dimensional Euclidean field theories:
\begin{equation}
    \langle \Phi_1(\x_1)\cdots \Phi_n(\x_n)\rangle \rightarrow \langle\Phi_1(\x_1)\star\cdots\star \Phi_n(\x_n) \rangle
    ,
\end{equation}
where $\x = (\tilde{x},x)$ and the $\Phi_{a_i}(\x_i)$ are some arbitrary operators. It will help to look at this correlation function after making partial Fourier transformations. First in the $x$-space, following \eqref{Moyal1},
\begin{align}
    &
    \big\langle\Phi_{a_1}(p_1,\tilde{x}_1) \star \cdots \star \Phi_{a_n}(p_n,\tilde{x}_n) \big\rangle
    \nonumber \\
    &= \bigg\langle \Phi_{a_1} \Big({\textstyle p_1,\tilde{x}_1-\kappa\sum^{n}_{j=2}p_j} \Big) \, \Phi_{a_2} \Big({\textstyle p_{2},\tilde{x}_2-\kappa\big(\!\sum^n_{k=3}p_{k}-p_1\big)}\Big) \, \cdot
    \nonumber \\ & \hspace{3em} \cdot
    \Phi_{a_3} \Big({\textstyle p_{3},\tilde{x}_3-\kappa\big(\! \sum^n_{k=4}p_{k} - \sum^2_{k=1}p_k \big)}\Big)
    \cdots \Phi_{a_n}\Big({\textstyle p_n, \tilde{x}_n + \kappa \sum^{n-1}_{k=1} p_k}\Big)\bigg\rangle
\end{align}
We can summarise this as 
\begin{equation}
    \big\langle \Phi_{a_i}(p_1,\tilde{x}_1) \star \cdots \star \Phi_{a_n}(p_n,\tilde{x}_n) \big\rangle = 
    \left\langle \prod^n_{i=1} \Phi_{a_i}\big(p_i,\tilde{x}_i-\kappa(J_{>}-J_{<})_i\big) \right\rangle
\end{equation}
subject to momentum conservation $\sum^n_{i=1}p_i=0$. Similarly, we can do a partial Fourier transform in the other direction and find
\begin{equation}
    \big\langle \Phi_{a_i}(x_1,\tilde{p}_1) \star\cdots\star \Phi_{a_n}(x_n,\tilde{p}_n) \big\rangle = \left\langle \prod^n_{i=1} \Phi_{a_i} \big(x_i+\kappa (\mathcal{E}_{>}-\mathcal{E}_{<})_i,\tilde{p}_i\big) \right\rangle
\end{equation}
Here we denote
\begin{equation}
    \left(J_{>}-J_{<}\right)_i = \sum^{n}_{k=i+1} p_k- \sum^{i-1}_{k=1}p_{k}
\end{equation}
and
\begin{equation}
    \left(\mathcal{E}_{>}-\mathcal{E}_{<}\right)_i = \sum^{n}_{k=i+1} \tilde{p}_k- \sum^{i-1}_{k=1}\tilde{p}_{k}
    .
\end{equation}
They denote the integrated momentum and energy density to the right of the operator inserted at $\x_i$. More precisely, they are defined through the stress tensor as
\begin{equation}
    2 \pi\left(\int_{x^i+\varepsilon}^{\infty}-\int_{-\infty}^{x^i-\varepsilon}\right) T_{00}\big(x^{1\prime}, 0\big) d x^{\prime 1} =  \left(\mathcal{E}_{>}-\mathcal{E}_{<}\right)_i
\end{equation}
and 
\begin{equation}
    2 \pi\left(\int_{x^i+\varepsilon}^{\infty}-\int_{-\infty}^{x^i-\varepsilon}\right) T_{01}\big(x^{1\prime}, 0\big) d x^{\prime 1} =  \left(J_{>}-J_{<}\right)_i
\end{equation}
if we take the $\tilde{x}=0$ surface as being space. 

Then, taking the derivative of the deformed correlation functions with respect to the deformation parameter $\kappa$, we obtain
\begin{align}
    &
    \partial_{\kappa} \langle \Phi_{a_1}(\x_1) \star\cdots\star \Phi_{a_n}(\x_n) \rangle
    \nonumber \\
    &= \sum^n_{i=1} 2\pi \epsilon_{\mu\nu} \epsilon^{\rho\sigma} \left(\int_{x^i+\varepsilon}^{\infty}-\int_{-\infty}^{x^i-\varepsilon}\right) d \x_\sigma^{\prime} T_\rho^{\mu}\left(\x^{\prime}+\varepsilon\right) \partial_{\x^\nu}\bigg\langle\Phi_{a_1}(\x_1) \star\cdots\star \Phi_{a_n}(\x_n) \bigg\rangle
    .
\end{align}
This is exactly the formula derived in \cite{Cardy:2019qao} and reproduced in \cite{Kruthoff:2020hsi} for the $T\bar{T}$ deformation of correlation functions in general Euclidean field theories! We recovered it here as a simple consequence of the matrix reformulation of the theory. 

\section{Deforming the actions}
So far, our treatment has focused on the deformation of observables like the $S$-matrix elements and correlation functions. It is also interesting to see how to track these prescriptions back to the level of the classical action, when such a notion is available. In this section, we will start by describing how to write the action of the two-dimensional free scalar field so as to reproduce the deformed $S$-matrix we describe above. Then we will look at a peculiar Lagrangian which describes the self-dual gravity in 4D where the interactions of the theory are effectively two-dimensional. We will then show that our matrix prescription can in fact be used to define the $T\bar{T}$ deformation of this model along the directions where the interactions occur, and furthermore, we will recognise this deformed theory to be the so-called Moyal deformed self-dual gravity theory. 

\subsection{The action for $T\bar{T}$ deformed free field theory}

In our prescription for deforming the $S$-matrix, we showed that we only needed to replace the free field operator with a matrix-valued field operator so that the regular product is replaced by the star product. 
The original action
\begin{equation}
S = \int \textrm{d}^2x\,\phi\square\phi
\label{eq:f_act}
\end{equation}
is now replaced by 
\begin{equation}
S_{T\bar{T}} = \textrm{Tr}\bigg(\eta^{\mu\nu}\phi\left[Q_\mu,\left[Q_\nu,\phi\right]\right]\bigg)
.
\label{eq:def_f_act}
\end{equation}
The fields now being matrices means that the action involves a trace operation and the kinetic term is replaced by a double commutator with the matrices $Q_0 \equiv Q$ and $Q_1 \equiv \tilde{Q}$. The commutator-action of these matrices define the momentum eigenvalues,
\begin{equation}
[Q_0,\phi] = p\phi,\,\,[Q_1,\phi] = \tilde{p}\phi,
\end{equation}
i.e. $[Q_\mu,\phi](p) = p_\mu \phi(p)$, where $\phi$ is an infinite matrix and $\phi(p)$ is its degree of freedom at $p$ (Section \ref{sec:MatGeo}). Now, the action \eqref{eq:def_f_act} is still quadratic in its fields, and the Green's function we extract from this action is exactly the same as the one we would obtain from it's un-deformed counterpart \eqref{eq:f_act}. However, the key difference lies in the way in which Wick's theorem applies to the products of field operators, as explained in the previous section. 

This remedial example gives us the recipe for deforming the classical action of a general two-dimensional theory: We write the kinetic term of the matrix theory just as \eqref{eq:def_f_act} and we write a trace of a product of field matrices replacing every interaction term. 

To exhibit the flexibility of our prescription, we will take a look at a non-covariant four-dimensional model whose interactions occur only along two dimensions. This theory describes the self-dual part of four-dimensional general relativity \cite{Chalmers:1996rq}.

\subsection{Self-dual gravity}

The linearized action for self-dual gravity can be written in terms of the fields $(\phi,\bar{\phi})$ which are related to the left and right-handed components of the graviton fluctuation through the relation:
\begin{equation}
(h,\bar{h}) = \left(\partial^2_u\phi, \partial^{-2}_u\bar{\phi}\right).
\end{equation}
Here $u$ is a light-cone coordinate: $(u,v,w,\bar{w}) = (t+z,t-z,x+iy,x-iy)$. In terms of these fields, the action for self-dual gravity reads (see also \cite{Smolin:1992wj, Bittleston:2022nfr})
\begin{align}
\label{SDG1}
    S_{\text{sdg}}[\phi, \bar{\phi}] = \int \dd^4x \, \bar{\phi} \left(\square \phi + \{\partial_u \phi, \partial_w \phi\}^{\text{Poisson}}_{u\text{-}w}\right),
\end{align}
where the Poisson bracket is defined as
\begin{equation}
\{\cdot,\cdot\}^{\text{Poisson}}_{u\text{-}w} = \partial_{u}(\cdot)\partial_{w}(\cdot) -\partial_{w}(\cdot)\partial_{u}(\cdot).
\end{equation}
The variation of \eqref{SDG1} returns Plebanski's second heavenly equation \cite{PlebanskiHeavenly} for the 4D graviton. The Moyal deformation of the self-dual gravity action was recently studied in \cite{Monteiro:2022lwm, Bu:2022iak},
\begin{align}
\label{SDG2}
    S_{\text{m-sdg}}[\phi, \bar{\phi}] = \int \dd^4x \, \bar{\phi} \left(\square \phi + \{\partial_u \phi, \partial_w \phi\}^{\text{Moyal}}_{u\text{-}w}\right)
    ,
\end{align}
where the Moyal bracket is defined as
\begin{align}
    \{A, B\}^{\text{Moyal}}_{u-w} \equiv \frac{1}{2i\kappa} \, [A,B] = \frac{1}{\kappa} \, A(u, w) \sin(\kappa (\cev{\partial}_u \vec{\partial}_{w} - \cev{\partial}_{w} \vec{\partial}_u)) B(u, w)
    \;.
\end{align}
Here, we will write \eqref{SDG2} as a matrix model, and argue that it is in fact the $T\bar{T}$ deformation of self-dual gravity.

\subsection{Matrix version of Moyal deformed SDG}

We make a Wick rotation on the $y$-coordinate to get signature $(2,2)$ and treat $w,\bar{w}$ as real coordinates. We rename the coordinates more suggestively as $(u,v,w,\bar{w}) = (x_1, x_2, \tilde{x}_1, \tilde{x}_3)$.

The action \eqref{SDG2} describes a 4D theory, but the Moyal bracket in its interaction term lives only on 2D. In a generic matrix model, for every momentum direction set by the $Q$-commutator eigenvalues, there is a dual direction set by the Fourier conjugate of $Q$-anticommutator eigenvalues, such that these dual directions generate a Moyal-type interaction \eqref{Moyal2} for the fields. The action \eqref{SDG2} lacks a dual direction for $v$ and $\bar{w}$. Therefore, we can only write it as a constrained matrix model \cite{Yargic:2022ycw}. We introduce
\begin{align}\begin{split}
    Q_1 = Q \otimes \mathbbm{1} \otimes \mathbbm{1}
    \;, \qquad
    Q_2 = \mathbbm{1} \otimes Q \otimes \mathbbm{1}
    \;, \qquad
    Q_3 = \mathbbm{1} \otimes \mathbbm{1} \otimes Q
    \;, \\
    \tilde{Q}_1 = \tilde{Q} \otimes \mathbbm{1} \otimes \mathbbm{1}
    \;, \qquad
    \tilde{Q}_2 = \mathbbm{1} \otimes \tilde{Q} \otimes \mathbbm{1}
    \;, \qquad
    \tilde{Q}_3 = \mathbbm{1} \otimes \mathbbm{1} \otimes \tilde{Q}
    \;,
\end{split}\end{align}
so that $[Q_i, Q_j] = [\tilde{Q}_i, \tilde{Q}_j] = 0$ and $[Q_i, \tilde{Q}_j] = \frac{i}{2\kappa} \delta_{ij} \mathbbm{1}^{\otimes 3}$. Let $\Phi$ and $\bar{\Phi}$ be hermitian matrices which are constrained as
\begin{align}
    [\tilde{Q}_2, \Phi] = [\tilde{Q}_2, \bar{\Phi}] = 0
    \;, \qquad
    [Q_3, \Phi] = [Q_3, \bar{\Phi}] = 0
    .
\end{align}
When represented in the common eigenbasis of $Q_i$, these constraints imply that $\Phi$, $\bar{\Phi}$ are Toeplitz in the 2-direction, and diagonal in the 3-direction. Then, we can write the Moyal-deformed self-dual gravity action \eqref{SDG2} as a constrained matrix model in 6 dimensions by
\begin{align}
    S_{\text{m-sdg}}[\Phi, \bar{\Phi}] = \Tr(\bar{\Phi} \left([Q_1, [Q_2, \Phi]] - [\tilde{Q}_1, [\tilde{Q}_3, \Phi]] + \tfrac{i}{2\kappa} \big[[Q_1, \Phi], [\tilde{Q}_1, \Phi]\big]\right))
    \;.
\end{align}
From the case we laid out in the bulk of this article, this matrix version of the self-dual gravity action \emph{is} the $T\bar{T}$ deformation of the theory described by the Chalmers-Siegel action! 

The scattering amplitudes of self-dual gravity are dressed in exactly the way as those of a two-dimensional theory deformed by $T\bar{T}$ would be. Let's take a look at the 3-point vertex, i.e., the tree-level $S$-matrix element of the Chalmers-Siegel theory,
\begin{equation}
    V(\p_1,\p_2) = \left(\p_{1}\times  \p_2\right)^2.
\end{equation}
This structure comes from the interaction term 
\begin{equation}
   \int d^4x \, \bar{\phi} \, \{\partial_u\phi,\partial_{w}\phi\}^{\text{Poisson}}_{u\text{-}w}
\end{equation}
in the action. Then, replacing the regular product with the matrix product, this term is replaced by 
\begin{equation}
    \tfrac{i}{2\kappa} \Tr(\bar{\Phi} \, [[Q_1,\Phi],[\tilde{Q}_1,\Phi]])
    ,
\end{equation}
which leads to a dressing of the vertex by CDD-like factors in the two-dimensional $u$-$w$ plane,
\begin{equation}
    V(\p_1,\p_2)\rightarrow V^{M}(\p_1,\p_2)=  \frac{(\p_{1}\times  \p_2)}{2\kappa} \left(e^{i\kappa(\p_1\times \p_2)}-e^{-i\kappa(\p_1\times \p_2)}\right)
    .
\end{equation}
Here the momenta are all along the $u$-$w$ plane. The difference between the two phases is a consequence of the commutator structure of the interaction.

\subsection{\texorpdfstring{From Heisenberg to $W_\wedge$ algebra}{From Heisenberg to wedge-W algebra}}

In this subsection we will study the symmetries of the Moyal deformed self-dual gravity action. Note that the relevant symmetry transformations are those that preserve the Poisson structure in the interaction term. 
The Moyal bracket is a deformation of the Poisson bracket, and it satisfies the Jacobi identity. The diffeomorphisms which preserve the Poisson structure of a 2D space are generated by Hamiltonian vector fields. Following \cite{Monteiro:2022lwm}, we will consider a basis of plane waves for the Hamiltonian functions.

For any $\alpha, \beta \in \mathbb{R}$, the matrix $e^{i\kappa (\beta Q - \alpha \tilde{Q})}$ has entries that form a plane wave when written in the $(x,\tilde{x})$-parameters \cite{Banburski:2022gcj},
\begin{align}
    \big(e^{i\kappa (\beta Q - \alpha \tilde{Q})}\big)(x, \tilde{x}) = e^{i\alpha x + i \beta \tilde{x}}
    .
\end{align}
We define $Q' \equiv - \kappa Q$ and $\tilde{Q}' \equiv \kappa \tilde{Q}$, and expand this matrix,
\begin{align}
    e^{-i (\beta Q' + \alpha \tilde{Q}')} = \sum_{a,b = 0}^{\infty} \frac{(i\alpha)^a}{a!} \frac{(i\beta)^b}{b!} \, \mathfrak{e}_{a,b}
    \;, \qquad
    \mathfrak{e}_{a,b} \equiv (\tilde{Q}'^a Q'^b)_{\mathrm{sym.}}
    ,
\end{align}
where $\mathfrak{e}_{a,b}$ is a matrix with the symmetrized ordering of its factors $Q'$ and $\tilde{Q}'$, for example, $\mathfrak{e}_{1,2} = (Q' \tilde{Q}'^2 + \tilde{Q}' Q' \tilde{Q}' + \tilde{Q}' Q'^2) / 3$ etc. These matrices satisfy the algebra
\begin{align}
\label{Walgebra}
    &\{\mathfrak{e}_{a,b}, \mathfrak{e}_{c,d}\}^{\text{Moyal}}
    \nonumber \\
    &= \sum_{s \geq 0} \frac{(-1)^s \kappa^{2s}}{4^{2s+1} (2s+1)!} \sum_{j=0}^{2s+1} (-1)^j \binom{2s+1}{j} [a]_{2s+1-j} [b]_j [c]_j [d]_{2s+1-j} \mathfrak{e}_{a+c-1-2s,b+d-1-2s}
    ,
\end{align}
where $[a]_n = a!/(a-n)!$ is the descending Pochhammer symbol. Note that $\tilde{Q}'$ and $Q'$ satisfy the Heisenberg algebra $[\tilde{Q}', Q'] = i \kappa / 2$, whereas \eqref{Walgebra} is the $W_\wedge$ algebra of Poisson-preserving diffeomorphisms \cite{Monteiro:2022lwm}. We showed here through the matrix formalism that the $W_\wedge$ algebra follows from symmetrized monomials of Heisenberg matrices.

The $W_\wedge$ algebra itself is a deformation of the $w_\wedge$ algebra, which is its limit $\kappa \rightarrow 0$,
\begin{align}
    \{\mathfrak{e}_{a,b}, \mathfrak{e}_{c,d}\}^{\text{Poisson}} = (ad - bc) \, \mathfrak{e}_{a+c-1,b+d-1}
    .
\end{align}
For a theory whose interactions follow a Poisson structure on a 2D space, the deformation of this Poisson bracket into a Moyal bracket is precisely the $T\bar{T}$ deformation of the theory. This is particularly the case for self-dual gravity.

\section{Outlook}

We have shown that replacing the spacetime dependence of a two-dimensional field theory with matrix index dependence leads to a very interesting set of non-localizing effects that are equivalent to those generated by the $T\bar{T}$ deformation. Our notion of deformation of Lagrangian theories differs from that of say \cite{Bonelli:2018kik} where instead of finding a Lagrangian written in terms of the original fields that solves the recursive $T\bar{T}$ flow equation, we replace fields as functions of space by matrices.
It would help to have a proper argument for why these two procedures should lead to the same results. There are some effects observed in the deformed Lagrangian that we cannot easily replicate in our proposal. For instance, it was argued in \cite{Chakrabarti:2020pxr} that tuning the $T\bar{T}$ deformation of a free boson tuned to an infinite value of the deformation parameter should leave us with the Lagrangian for a chiral boson. This is something we cannot straightforwardly see from the CDD phase. 

This notion is very much in keeping with the various gravitational dressing perspectives \cite{Dubovsky:2018bmo,Dubovsky:2017cnj,Cardy:2018sdv,Tolley:2019nmm,Mazenc:2019cfg,Caputa_2021}, or those involving dynamical changes of coordinates \cite{Conti:2018tca,Guica:2019nzm,Coleman:2019dvf}, yet somehow it is even closer to the expectation that the $T\bar{T}$ deformation does something to the states of the original theory very much akin to what a tensor network representation of such a state would do \cite{Kruthoff:2020hsi,Caputa:2020fbc}. It would be very interesting to pursue this connection. 

An object that we haven't touched upon at all in this article is the $T\bar{T}$
deformed partition function \cite{Datta:2018thy,Aharony_2019}. In order to see how the energy levels are deformed, we relied on the Bethe equations of integrable models. An important next step for solidifying our proposed connection would be to compute the partition function and the deformed energy levels for non-integrable theories. 

An important generalization of the $T\bar{T}$ deformation is the so called $T\bar{T}+\Lambda_2$ deformation \cite{dSdSTTbar,Lewkowycz:2019xse}. It has recently played a key role in understanding the microstates of the cosmic horizon in three-dimensional dS space \cite{Coleman:2021nor,Shyam:2021ciy}. Capturing the additional $\Lambda_2$ deformation in our approach would provide a matrix representation for said de Sitter microstates! 

Among other notable cousins of the $T\bar{T}$ deformation is the so called $J\bar{T}$ deformation \cite{Guica:2017lia}. This deformation also has a geometric interpretation \cite{Aguilera-Damia:2019tpe,Tolley:2019nmm} and perhaps it has a matrix realisation. It would be interesting to see if such a deformation could be applied to self-dual Yang-Mills theory. 

It is also worth mentioning \cite{Guevara:2022qnm} where a color version of the clock and shift matrices featured above were studied, and were used to construct gravitational scattering amplitudes. It would be very interesting to see if we can make contact with that construction.

\section*{Acknowledgements}

Microsoft supported this research both by funding researchers and providing computational, logistical and other general resources.

We are grateful to Jaron Lanier and Kevin Scott of Microsoft in particular for support of this project. We also thank Lee Smolin for discussions.

V.S.~is supported by the Branco Weiss Fellowship - Society in Science, administered by the ETH Zurich.

\bibliographystyle{utphys}
\bibliography{References}

\end{document}